# Quantum effects on the Peierls stress for screw dislocations in α-Fe

Jiaqi Sun[1], Yang Li[1], Boyang Gu[1], David McDowell[2,3], Youping Chen[1]
[1]Department of Mechanical and Aerospace Engineering, University of Florida, Gainesville, FL 32611, USA
[2]Woodruff School of Mechanical Engineering, Georgia Institute of Technology, Atlanta, GA 30332, USA
[3]School of Materials Science and Engineering, Georgia Institute of Technology, Atlanta, GA 30332, USA

## Abstract

The large discrepancy between the Peierls stress calculated using atomistic methods and that measured in experimental studies is well known. Despite significant research efforts over the past decades, our understanding of this mismatch remains limited. This work presents a multiscale study on the kinetics of screw dislocations in α-Fe at 0.7K, employing the phonon description of temperature and using the Concurrent Atomistic-continuum (CAC) method as the simulation tool. Our simulations are shown to preserve the Bose-Einstein distribution of phonon modes and reproduce the nucleation and propagation of kink-pairs. Large-scale CAC simulations with dislocation line length ranging from 0.1 to 2.1μm —a typical range of lengths for screw dislocation lines—are performed, providing a quantitative and mechanistic understanding of the phonon-dislocation in α-Fe. The experimentally measured Peierls stress in α-Fe has been successfully reproduced, and the quantum effects on Peierls stress are quantified. Dependencies of phonon-dislocation interactions on dislocation line length and specimen size are explored and discussed.

## 1. Introduction

Plastic deformation in crystalline materials, including metals and semiconductors, occurs primarily through the motion of dislocations. The intrinsic lattice resistance of a crystal to dislocation motion is characterized by the Peierls stress, which is defined as the minimum stress required to move a dislocation through a crystal lattice at zero temperature. This stress is a primary determinant of dislocation mobility. It is also a key controlling parameter in mesoscale models such as crystal plasticity and discrete dislocation dynamics, determining the threshold stress for dislocation movement. A long-standing problem universally encountered in the study of plastic deformation in body-centered cubic (BCC) crystals is the large discrepancy between the Peierls stress computed using atomistic methods and that obtained via experimental measurement. For instance, the Peierls stress in α-iron measured in experiments is about 370-390 MPa[1-5], whereas that from classical atomistic simulations[6,7] and *ab initio* calculations[8-10] ranges from 1.0–1.4 GPa. Similar discrepancies have been reported for many other BCC crystals[11].

Various theories have been proposed to explain this discrepancy[5,12]. Recent efforts to understand this discrepancy have focused on quantum effect. Based on an analysis using quantum transition state theory, Proville et al.[11] suggested that the discrepancy is due to the effect of "crystal zero point vibrations", a quantum effect that cannot be captured by classical theories or atomistic simulations. This conclusion was later challenged by Freitas et al.[13] who computed the resistance to dislocation motion in α-Fe using classical Molecular Dynamics (MD) and the Ring-Polymer MD (a quantum dynamical simulation method), respectively, and compared the results in the limit of zero temperature. Their computational results suggest that the effect of "the zero-point vibrations" are noticeable for T< 50 K but account for only a modest (≈13%) overall reduction in the Peierls barrier[13]. Accordingly, they concluded that "the long-standing discrepancy between calculated and experimental estimated Peierls stress in BCC metals remains and further work is needed to settle this outstanding issue in physical metallurgy"[13].

It is noted that all the experiments that measured flow stress at the specimen level[1] or local stress at the individual dislocation level[14] were conducted at low but finite temperature for dislocations with a finite line length[5,14-16]. By contrast, most atomistic methods, including Density Functional Theory (DFT) calculations and MD simulations "were concerned with straight infinite dislocations at 0K temperature"[17]. This is because the length and time scales involved in the kinetics of screw dislocations in BCC crystals at low temperature are too large to be modeled by methods with nanoscale resolution such as MD[18], which is currently the only established tool that can simulate the kinetic processes of dislocations without the need to assume the underlying mechanism. The stress discrepancy thus represents not only "a major difficulty in physics of dislocations"[5], as stated by Caillard[5], but also a significant challenge to computational methods for understanding of kinetic processes of dislocations; it therefore limits our ability to connect theoretical models and experimental measurements of fundamental mechanical properties and processes of defect migration in crystalline materials.

This work aims to address this challenge by performing mesoscale simulations and employing phonon description of temperature, using the Concurrent Atomistic Continuum (CAC) method[19]. The CAC method is a coarse-grained atomistic method that reduces the degrees of freedom (DOFs) of an atomistic system by utilizing the continuous structure and dynamics of crystals at the lattice level, while preserving the discrete atomic structure and motion within each lattice cell[20-22]. The CAC formulation is derived bottom-up from the atomistics and is otherwise identical to the underlying atomistic model when discretized at the primitive unit cell level[22,23]. The theoretical formulation unifies not only atomistic and continuum representations of conservation laws and the definitions of stress and heat flux[22,24,25], but also the definition of temperature[23]. It thus realizes full applicability of MD in predicting nanoscale materials properties and behavior, including dislocation core structures, phonon dispersion relations, phonon scattering on dislocations, etc. Due to its non-local force-displacement relations, CAC can simulate the nucleation and propagation of dislocations and stacking faults in the coarse-grained finite element region through relative motion between elements, thereby addressing the length scale challenge in multiscale simulation of defect structures and dynamics. The accuracy and efficiency of CAC have been tested through one-to-one comparisons with MD in nano-scale simulations of crack nucleation and branching[26,27], phase transitions[28], dislocation nucleation and propagation[29-33], dislocation loop formation[34-37], dislocation interactions with interfaces and other defects[38-44], dislocation formation during epitaxial processes[45,46], as well as phonon-dislocation[47-49] and phonon-interface interactions[50,51].

The reminder of this paper is organized as follows. In Sec. 2, we introduce the methodology, the computer models and simulation setup, and calculation of Peierls stress for BCC α-Fe crystals. In Section 3, we present simulation results of the dynamics of dislocations at low temperature and quantify the effects of temperature definition, the dislocation length, and the specimen size on Peierls stress. This paper ends with a brief summary of major findings in Section 4.

## 2. Methodology

### 2.1 Phonon description of temperature

In classical mechanics and conventional molecular dynamics (MD) simulations, temperature is defined as the kinetic temperature in the molecular kinetic theory. The equipartition theorem of kinetic theory asserts that the average kinetic energy of every particle $k$ in an equilibrium gaseous system with temperature $T$ is $<\frac{1}{2}m\boldsymbol{v}_k^2>=\frac{3}{2}k_BT$ , thereby defining equilibrium temperature in terms of the mean kinetic energy of atoms.

Temperature in quantum theories is described in terms of phonons. Phonons are quantized vibrational modes of lattices and can be experimentally imaged[52,53]. The average number of phonons with wave vector $\boldsymbol{k}$ and branch $v$, $n(\boldsymbol{k}, v)$, is a function of temperature, $T$, that includes the Planck constant $\hbar$ and the Boltzmann constant $k_B$. This relationship is called Bose-Einstein distribution, which gives the kinetic energy $K$ of each phonon mode ($\boldsymbol{k}$, $v$) as

$$K(\boldsymbol{k},\nu)=\frac{1}{2}\hbar\omega(\boldsymbol{k},\nu)\left[\frac{1}{2}+n(\boldsymbol{k},\nu)\right], \text{ where } n(\boldsymbol{k},\nu)= n(\omega,T)=1/\left[\exp\left(\frac{\hbar\omega(\boldsymbol{k},\nu)}{k_BT}\right)-1\right]. \quad (1)$$

The difference between the descriptions of classical temperature and quantum temperature can be quantified by comparing a system's total kinetic energy of phonons with that of atoms[54]. Fig. 1 presents such a calculation for single crystal α-Fe. As can be seen from Eq. (1), at 0K quantum temperature, α-Fe has a zero-point energy equivalent to 142.7K (classical temperature). This means that there are corresponding motions at 0K quantum temperature, generally referred to as the zero-point vibrations.

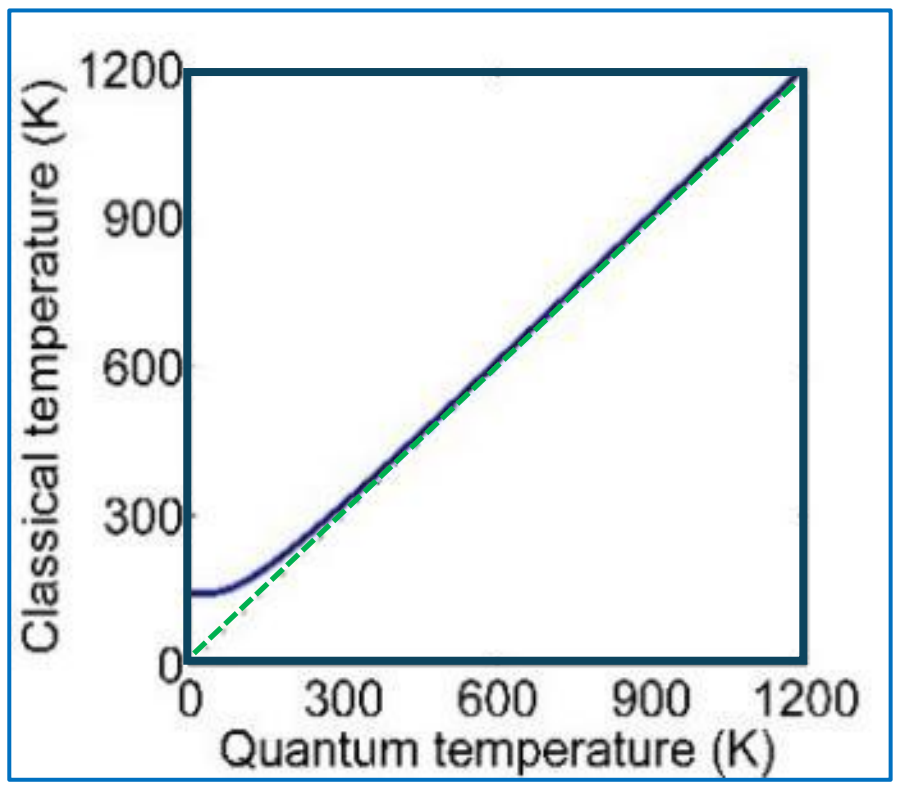

Fig. 1 Quantum vs classical temperature for α-Fe, showing that the two descriptions converge with increasing temperature, i.e., approaching the dotted line.

In Fig. 2, we plot the distribution of phonon modes based on Eq.(1). It is seen that the dominant phonon modes at both 100K and 0.7K are the low-frequency (i.e., long-wavelength) phonons. This indicates, in addition to the zero-point energy, there is a fundamental difference in the distribution of phonon modes between classical and quantum descriptions, with long wavelength phonons dominating at low temperature in quantum calculations, while classical temperature simulations assume a random distribution that ignores such shifts of the phone mode distribution at low temperature.

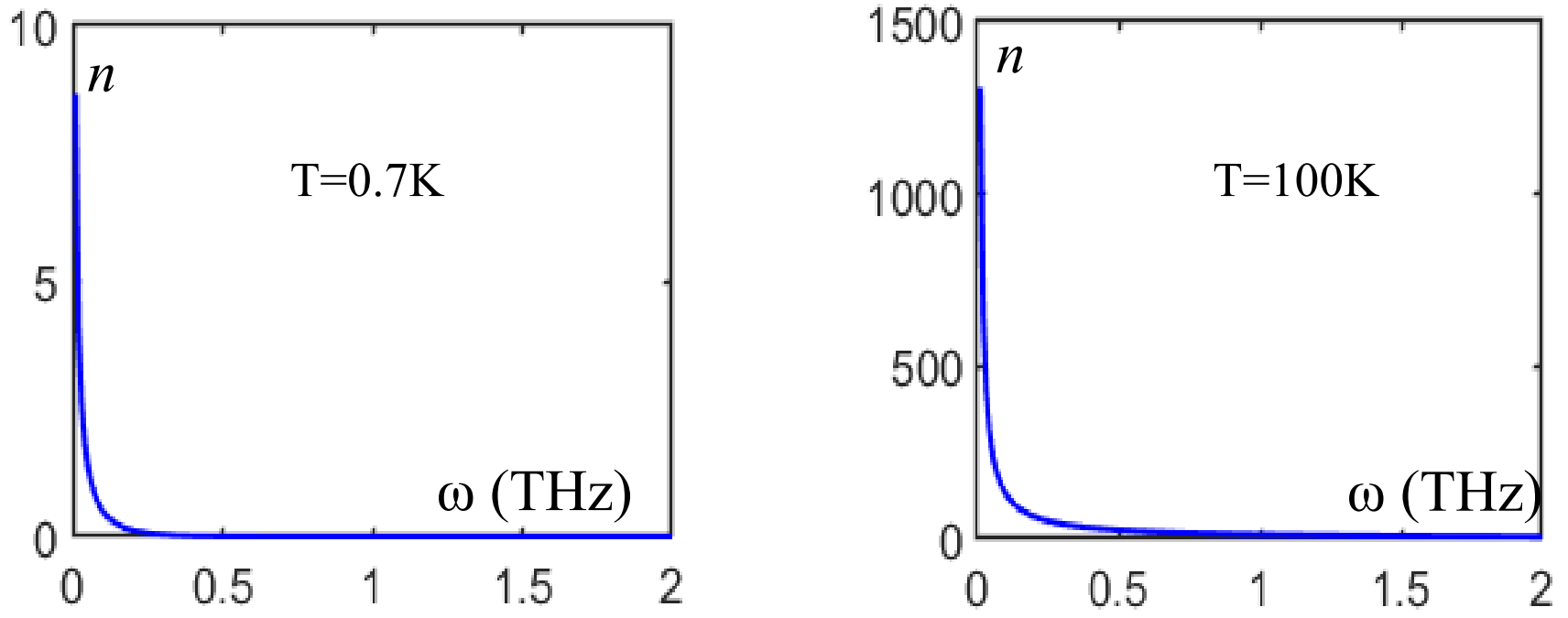

Fig.2 Average phonon number $n$ at T= 0.7 (left) and 100K (right) as a function of frequency ω for α-Fe.

Recall that phonon scattering by phonons, defects, interfaces, surface boundaries, etc., is the origin of thermal resistance or finite thermal conductivity. The fact that there is a very small number of phonons at T =0.7K, compared with that at 100K, as shown in Fig.2, indicates that the probability of phonon-phonon scattering decreases with decreasing temperature. Moreover, experimental studies[55] have shown that at low-temperature, thermal conductivity increases linearly with specimen size for T<30K, indicting that at low temperature phonon scattering by specimen boundaries constitutes the primary thermal resistance, while phonon-phonon scattering becomes negligible.

Since temperature is defined in the CAC method as a derived quantity, we can employ either the phonon definition of temperature or classical definition of temperature to quantify the effect of the difference between classical and quantum descriptions of temperature, i.e., the quantum effect, on computed

Peierls stress. This work performs CAC simulation at a very low temperature at which phonon-phonon scattering is negligible. To avoid phonon scattering by the surface boundaries, we simulate the system with a constant background temperature by generating phonons that obey the Bose-Einstein phonon distribution at one boundary of the specimen while absorbing phonons that reach the other boundaries to avoid phonon scattering by specimen boundaries. Without a strong phonon-phonon and phonon-boundary scattering to thermalize phonons, the Bose-Einstein phonon distribution can be preserved in a low-temperature simulation governed by Newton's second law. The quantum effect on dislocation motion and Peierls stress can thus be quantified.

## 2.2 Computer models and simulation setup

Figure 3 presents a schematic of the computer models for CAC finite element simulations of BCC iron single crystals containing one screw dislocation, including the dimensions of the models and the boundary and loading conditions. Each CAC computer model is discretized with uniformly meshed rhombohedral finite elements, with each element containing 16x16x16 primitive unit cells of BCC iron, equivalent to 4096 atoms per finite element, as shown in Fig. 4. The total number of degrees of freedom (DOFs) of the finite element models is about 1.95% of an equivalent fully resolved atomic model. A [111] screw dislocation is created by applying the corresponding Volterra displacement field at the center of the model, as shown in Fig. 4, through relative displacement of the finite elements. To simulate a finite length dislocation without artificial image force interactions, free boundary conditions are applied on all surfaces.

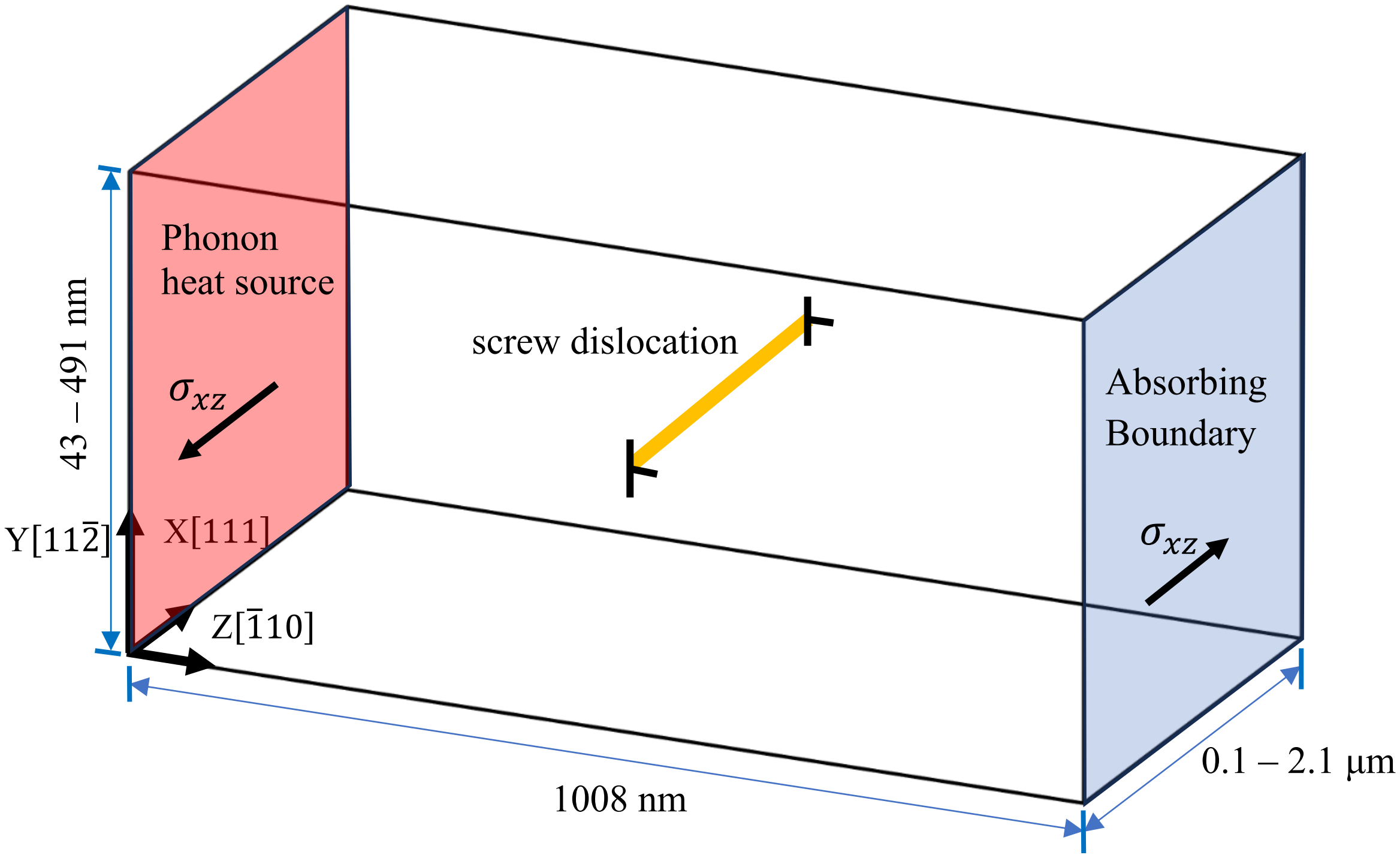

Fig. 3 Schematic of the computer models. The X axis corresponds to the dislocation line direction [111], Y axis to the dislocation motion direction [11$\bar{2}$], and Z axis to the phonon propagation direction [$\bar{1}$10]. The surface marked in red is the location of the heat source while blue the absorbing boundary.

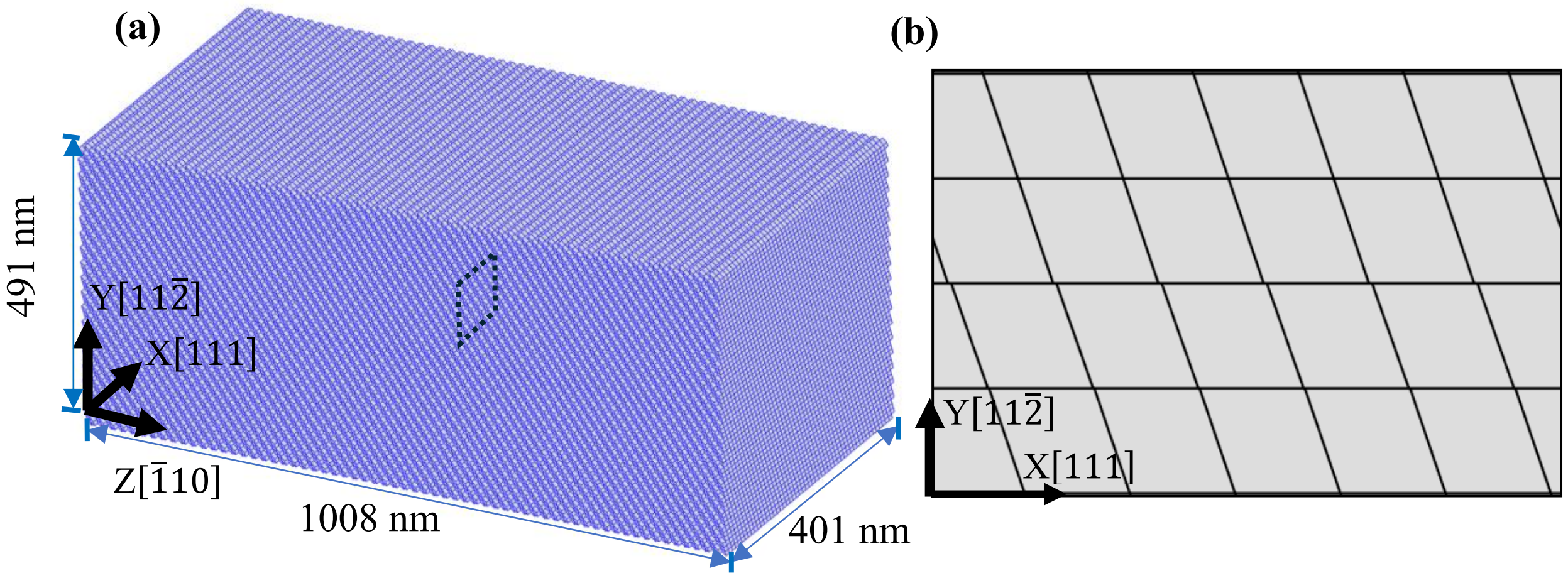

Fig. 4 (a) The finite element model of the specimen with dimensions of 401 nm x 491 nm x 1008 nm, (b) zoomed-in plain view of the boxed region shown at the center of the specimen in (a), showing a dislocation manifested by relative displacements between nodes of adjacent finite elements.

Table 1 summarizes the dimensions of the computer models. To quantify dislocation interaction with phonons with wavelength up to 1μm, the dimension of the specimen along the z axis has a fixed value of 1008 nm. To quantify the effect of dislocation length on Peierls stress, a set of computer models with a fixed thickness of 123 nm were constructed with specimen length ranging from 100 nm to 2104 nm and the dislocation line length ranging from 0.1μm to 2.1μm. The largest model in this set has dimensions 2104 nm x 123 nm x 1008 nm, corresponding to an atomically resolved model that contains 20.85 billion atoms. To investigate the effects of specimen thickness, an additional set of 6 computer models were constructed with thickness ranging from 43 nm to 491 nm. The largest model in this set has dimensions of 2104 nm x 245nm x1008 nm and contains 10.18 million finite elements, corresponding to 41.70 billion atoms in the equivalent atomically resolved MD model.

Table 1 Dimensions of the CAC finite element models and equivalent number of atoms

| | Dimensions of the Computer Model (nm) | Dislocation Line Length (nm) | Number of Finite Elements | Number of Equivalent Atoms |
|---|---|---|---|---|
| Set 1 | 100x123x1008 | 99.73 | 234,479 | 0.96 billion |
| | 201x123 x1008 | 200.6 | 478,289 | 1.96 billion |
| | 401x123x1008 | 400.6 | 963,200 | 3.94 billion |
| | 1006x123x1008 | 1005.9 | 2,495,020 | 10.21 billion |
| | 2104x123 x1008 | 2101.9 | 5,091,150 | 20.85 billion |
| Set 2 | 401x43x1008 | 400.6 | 327,600 | 1.34 billion |
| | 401x123x1008 | 400.6 | 963,200 | 3.94 billion |
| | 401x245x1008 | 400.6 | 1,926,400 | 7.89 billion |
| | 401x368x1008 | 400.6 | 2,889,600 | 11.84 billion |
| | 401x491x1008 | 400.6 | 3,852,800 | 15.78 billion |
| | 2104x43x1008 | 2101.9 | 1,697,050 | 6.95 billion |
| | 2104x245x1008 | 2101.9 | 10,182,300 | 41.70 billion |

The Proville-Rodney-Marinica EAM potential[11] for BCC iron is employed for all simulations reported in this work. This potential was optimized for modeling the core structure and mobility of screw dislocations in BCC iron. It accurately reproduces the compact, non-degenerate core configuration observed in ab initio (DFT) calculations. Accordingly, errors in interatomic potential resulting from the approximation of the profile of the energy barrier to form kink pairs are minimized.

To quantify the effect of the temperature description on Peierls stress, each computational model was simulated using two temperature representations: the kinetic temperature definition and the phonon description of temperature. For simulations using the kinetic description, a small damping force was added to the elements located in the absorbing boundary whenever the specimen temperature exceeded the target temperature of the simulation. For simulations using the phonon description of temperature, phonon representations of the heat source and the absorbing boundary condition (ABC) were implemented. The heat source, containing two layers of finite elements, with a thickness of 7.2 nm in the $[\bar{1}10]$ direction, was located at one end of the model. The absorbing boundary (heat sink) was located at the other end of the model, containing 6 layers of finite elements, the thickness of which is 21.6 nm in the $[\bar{1}10]$ direction.

The phonon heat source was constructed as a superposition of phonons, using the Bose-Einstein distribution, which links the phonon number, $n$, for each phonon mode (k, ν) to temperature, T, as shown in Eq. (1). The phonon frequencies in Eq. (1) were obtained through the phonon dispersion relations of the material system. In this work, the phonon dispersion relations of a CAC finite element (FE) model are obtained by solving the dynamic matrix of the CAC equations. The process starts with displacing finite element nodes with displacements of 0.001Å in each X, Y, and Z direction, respectively, to calculate the induced forces, leading to calculations of the force constants and the dynamical matrix, from which we proceed to solve the eigenvectors and frequencies to obtain the phonon dispersion curves. The atomic displacement within the heat source region is then calculated as

$$u_{i\mu\lambda}(\boldsymbol{r},t) = \sum_{j=1}^{N} A_j \varepsilon_{ji\mu\lambda} \exp\left[i\left(k_j(\boldsymbol{r}-r_0) - \omega_j t\right)\right] \tag{2}$$

where $u_{i\mu\lambda}$ is the is the $\mu$th component of the displacement of atom $i$, $\lambda$ is the phonon branch, $r$ is the position within the heating region, $N$ is the number of phonon modes, which is equivalent to the total degrees of freedom of the computer model; $\boldsymbol{\varepsilon}$ is the eigenvector; $r_0$ are the spatial center positions of the heating region; $k_j, \omega_j$ , $A_j$ are the central wavevector, frequency, and amplitude of phonon mode $j$; and the amplitude is calculated using an approximate formula in solid state physics[56]

$$A_j{}^2 = 4\left(n + \frac{1}{2}\right)\frac{\hbar}{\rho V \omega_j} \tag{3}$$

where $n$ is the phonon number, $\rho V$ is the mass of the system. During the simulation, the atoms in the heat source region are prescribed to follow this displacement function to model continuous excitations of phonon waves in time. The absorbing boundary at the other end of the specimen then absorbs all the waves that propagate from the heat source and reach it. This technique generates a kinetic energy distribution with a constant average in time and across the entire specimen with phonons that obey the Bose-Einstein distribution.

### 2.3 Verification

#### A. Equilibrium state with phonon description of temperature

Figures. 5 and 6 present the spatial and temporal distributions of kinetic energy and the x-component of the displacement field, respectively, for a dislocation-free specimen with dimensions of 401 nm x 123 nm x 1008 nm at 0.7K. The heat source was constructed based on 23 million phonons that may exist in the computer model. It is seen that the kinetic energy, if averaging over a time interval, is nearly uniform in space and time across the computer model with no significant phonon-boundary scattering. Meanwhile, the time sequence of the displacement fields in Fig. 6 clearly indicates a constant displacement field, upon a time-interval averaging. Both Fig. 5 and Fig. 6 indicate that the computer model has reached an equilibrium state, as both time-interval averaged kinetic energy and displacement field are a constant in space and time.

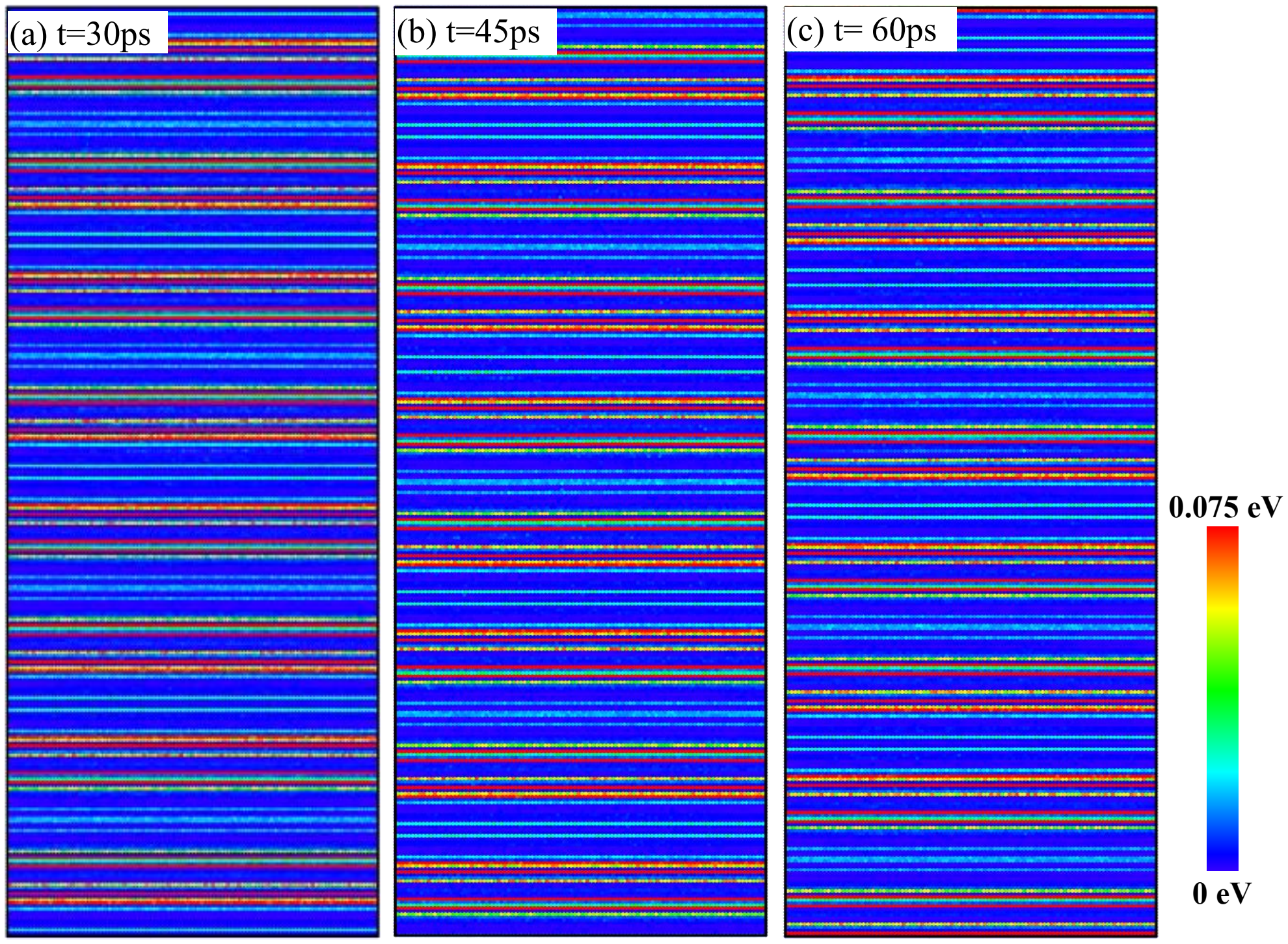

Fig. 5 Time sequence of the kinetic energy of a dislocation-free CAC model with dimensions of 401 nm x 123 nm x1008 nm, showing the equilibrium state at a phonon temperature of 0.7K.

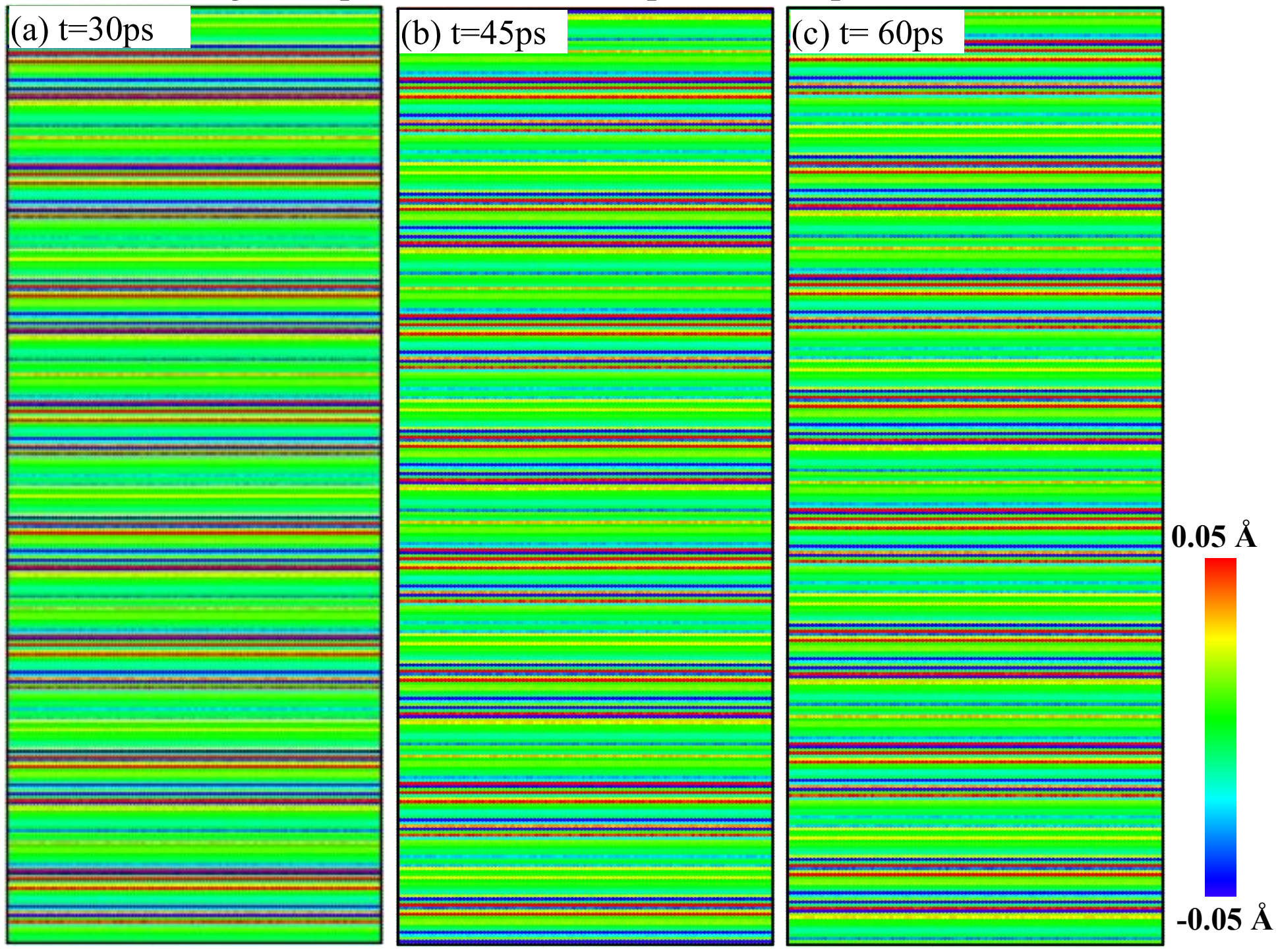

Fig. 6 Time sequence of [111]-direction displacement field in a dislocation-free CAC model with dimensions of 401 nm x123 nm x1008 nm at phonon description of temperature of 0.7K.

In Fig. 7, we plot the total kinetic energy of the specimen with a dislocation, as well as that of a small region containing the dislocation, as a function of time. It is seen from the plateaus of total kinetic energy in Fig. 7 that the system has reached an equilibrium state since the averaged kinetic energy for the entire specimen, as well as that for the region containing the dislocation, does not change with time. This result further demonstrates that the phonon representation of the heat source, together with the absorbing boundary, is working properly and has produced an equilibrium state. These simulation results demonstrate that the simulation methodology can preserve the Bose-Einstein distribution of phonons at low temperatures.

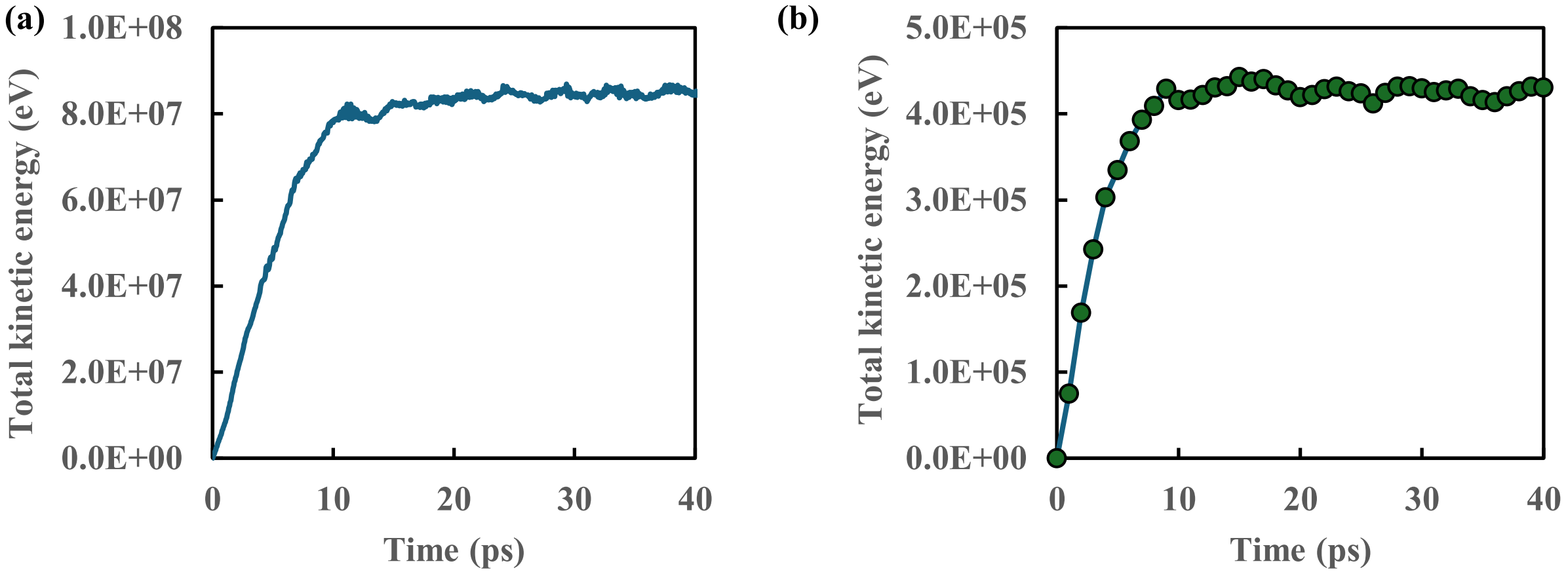

Fig. 7 Total kinetic energy as a function of time for (a) the entire computer model with dimensions 401 nm x123 nm x1008 nm, and (b) the region with dimensions 401 nm x123 nm x 50 nm containing the dislocation.

## B. Comparing Peierls stress, dislocation motion, and core structure by MD and CAC

To quantify the accuracy of CAC with respect to MD, we compare CAC and MD simulation results of Peierls stress, dislocation motion and core structure. To remain within the typical computational scale of MD simulations, we constructed small-scale MD and CAC models representing the same physical system and containing an equivalent number of atoms. The computer models measure 102.7 nm × 61.2 nm × 42.5 nm, corresponding to approximately 20 million atoms. The MD simulation resolves the entire system at full atomic resolution, whereas the CAC model consists of 4,933 finite elements. In this section, the kinetic temperature definition was employed in both CAC and MD simulation for comparison purposes.

The determination of Peierls stress through CAC and MD simulations was carried out in two steps. First, an incremental shear stress was applied in the XZ plane on the two surfaces of the model shown in Fig.3. The applied shear stress starts from the experimentally reported value of 375 MPa. Every 50 ps, the shear stress was increased by a fixed increment of 20 MPa. Dislocation motion was monitored using the Dislocation Extraction Algorithm (DXA), as well as the potential energy contours. The stress at the onset of dislocation motion was recorded. Second, a series of independent simulations were conducted under a constant applied shear stress to eliminate dynamic artifacts. The initial applied stress was set to the stress value recorded in the first step. If the dislocation exhibited steady, continuous glide under this stress, the applied stress was decreased by 0.5 MPa in the subsequent simulation. This process was repeated until continuous dislocation motion no longer occurred. The lowest applied stress that successfully sustained continuous glide was identified as the critical resolved shear stress (CRSS) at very low temperatures. In experimental studies, the Peierls stress estimated by extrapolating the CRSS to 0 K was very close to the CRSS measured at a near-zero temperature of 0.7K[1,3]. We thus take the CRSS measured at 0.7K in our simulations as the Peierls stress.

The values of the CRSS at 0.7K obtained in our simulations at kinetic temperature 0.7K were 1361.8 MPa from MD and 1382.9 MPa from CAC. This close agreement indicates that the CAC method can accurately reproduce the results of MD for Peierls stress, despite a significant reduction of the degrees of freedom (DOFs).

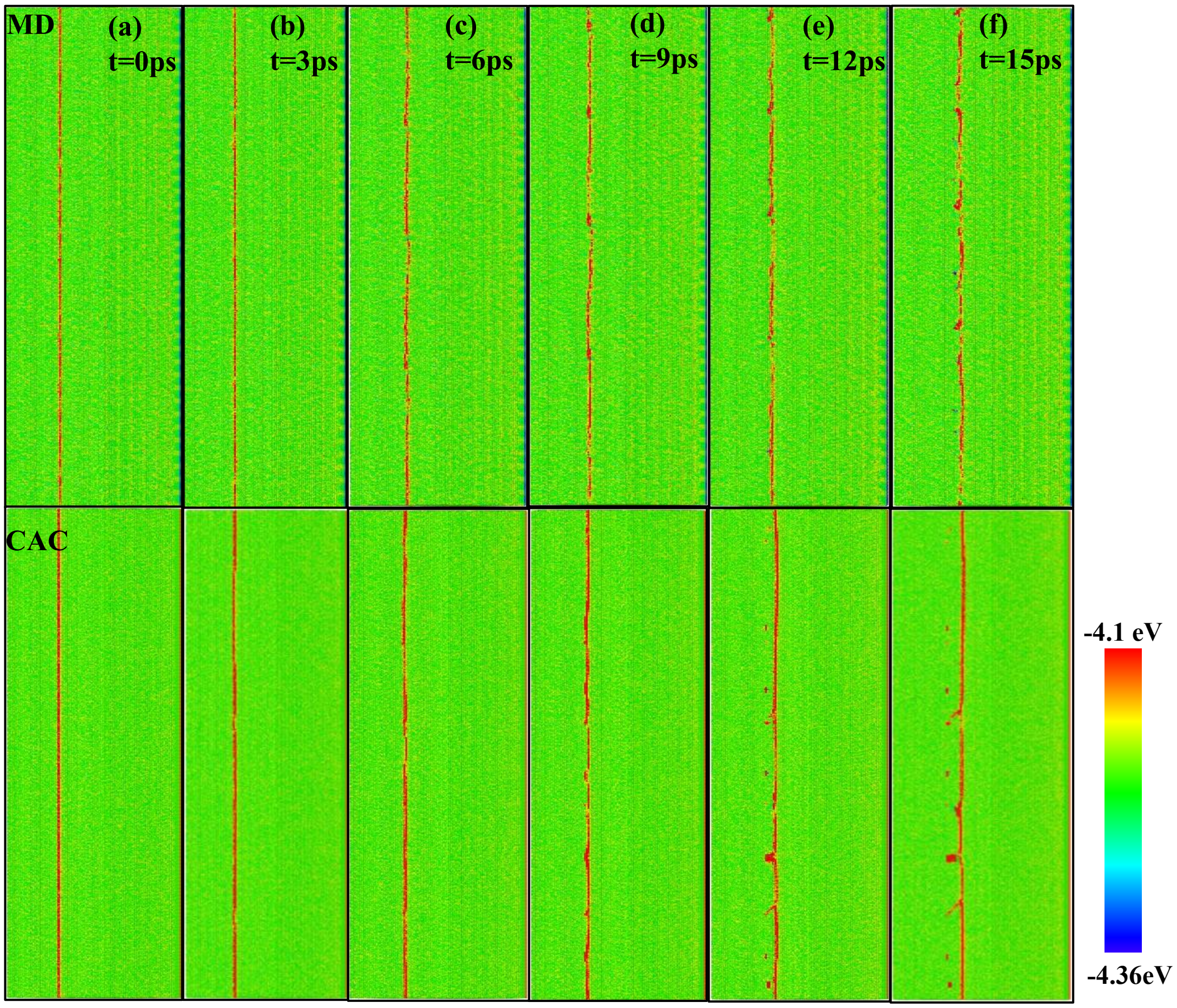

Fig. 8 Time sequence of potential energy density at 0.7K (kinetic temperature) obtained by MD (top row) and CAC (bottom row) for a computer model with dimensions of 102.7 nm x 61.2nm x 42.5 nm and containing a dislocation with length of 102.7 nm, showing the dislocation motion.

In Fig. 8 we present the time sequence of the potential energy obtained by CAC and MD respectively, to further show the motion of the dislocation. As can be seen from Fig.8, both MD and CAC have captured the continuous motion of the dislocation under an applied shear stress, demonstrating that the finite elements in CAC can capture the dislocation motion though the relative movements of elements and reproduce the time sequence of dislocation motion that is comparable with MD. Fig. 9 presents the DXA analysis results of the dislocation motion, as well as the differential displacement map of the dislocation core structure. These results further demonstrate that CAC, with a uniform coarse-scale finite elements, reproduces both the dislocation motion and the atomic-scale dislocation core structure, in good agreement with atomically resolved MD.

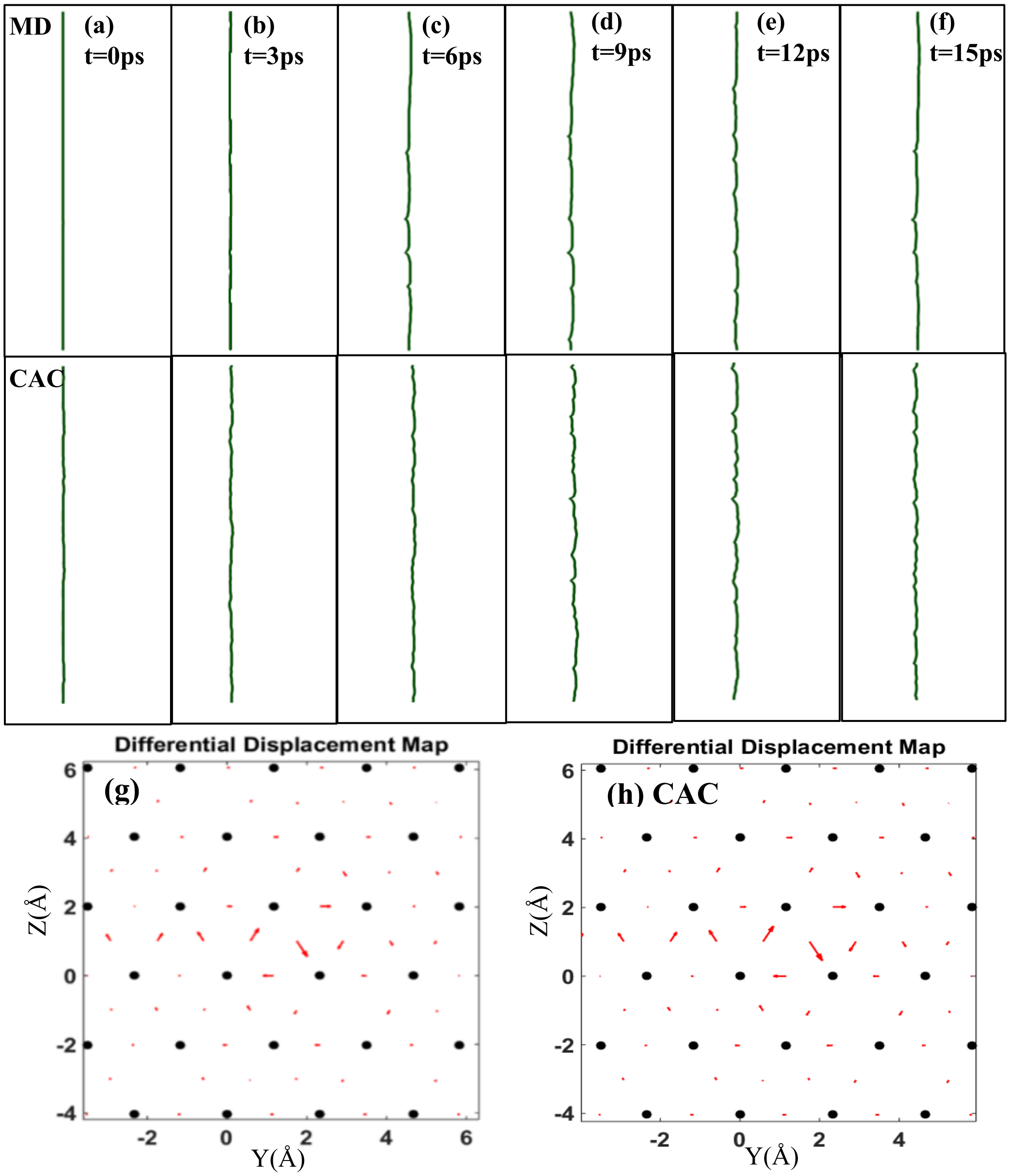

Fig. 9 Comparisons of MD and CAC simulation results for a dislocation line length of 100 nm at 0.7K kinetic temperature; (a-f) the DXA result of the time sequence of the dislocation motion; (g) and (h) differential displacement maps for the dislocation core structure obtained through relaxation at 0.7K.

### 2.4 Role of short wavelength phonons

CAC finite elements that employ the conventional tri-linear shape functions cut off phonons with wavelength smaller than the finite element size. These phonons can be added back to the shape functions directly[57] or through a reformulation of the conservation equation of linear momentum[23,58]. To quantify the role of the short wavelength phonons that are cut off by the tri-linear shape function of the finite element at 0.7K, In Fig. 10, we compare CAC results of the phonon density of states with and without the short wavelength phonons of a BCC iron crystal containing a dislocation. As can be seen from Fig.10, the CAC model with the short wavelength phonons being represented through the shape functions is capable of reproducing the full phonon spectrum as that by MD.

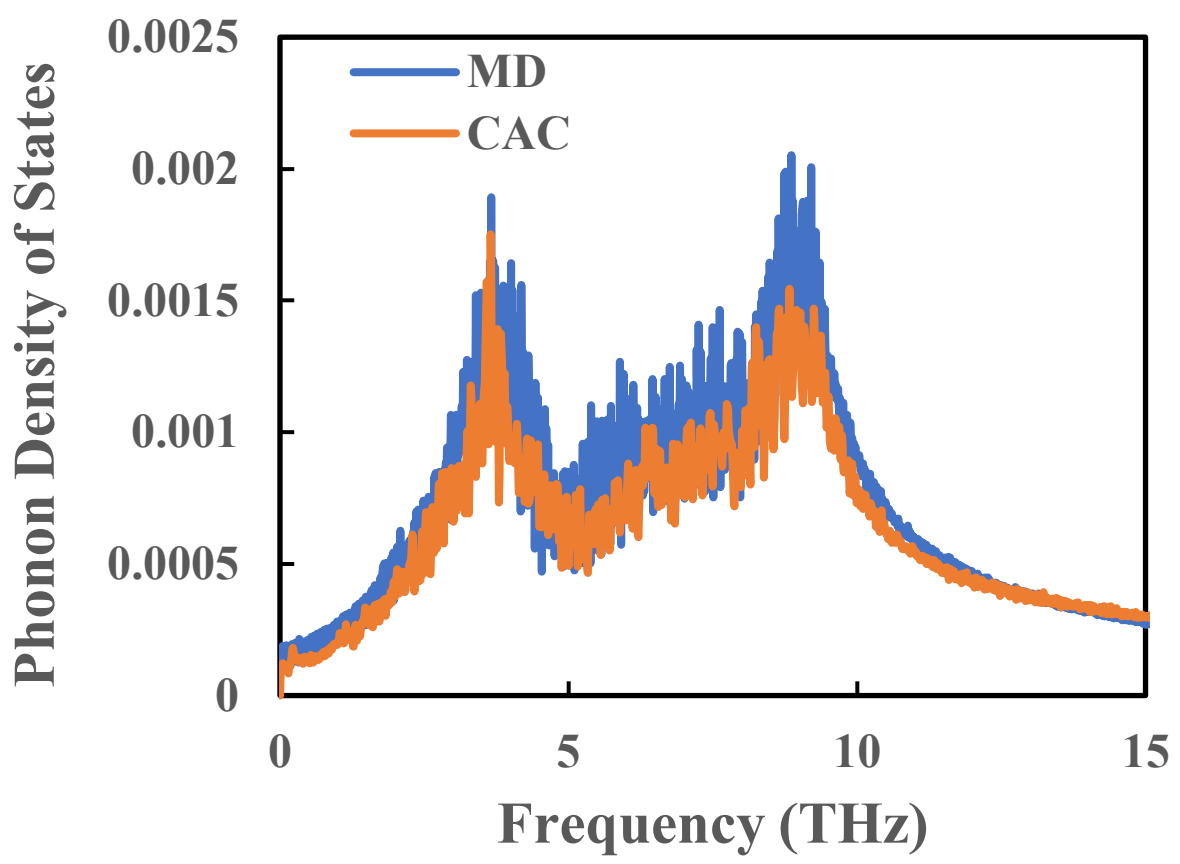

Fig. 10 Comparisons of MD and CAC simulation results of normalized phonon density of state (PDOS) at 0.7K (kinetic temperature), (a) without and (b) with short wavelength phonons.

Table 2 compares the values of Peierls stress obtained from CAC simulations with and without the short wavelength phonons being added to the shape functions, using the classical and phonon descriptions of temperature, respectively. As can be seen from Table 2, the difference between Peierls stress obtained with and with short wavelength phonons is less than 1% in both cases. Thus, the short wavelength phonons do not play a significant role at a low temperature of 0.7K. This is because (1) at low temperatures, the vibrational spectrum of a crystal is dominated by long-wavelength phonons, and (2) the interactions between long dislocations and very short wavelength phonons are weak and so the short wavelength phonons play no significant role on the motion of long dislocations. Furthermore, Table 2 shows that the phonon description of temperature accounts for the significant reduction of the Peierls stress.

Table 2 CAC simulation result of Peierls stress (MPa), with and without including the short wavelength phonons, in a specimen with dimension of 401 nm x123 nm x1008nm.

| | Without short wavelength phonons | With all phonons |
|---|---|---|
| Classical description of temperature | 1222.3 | 1220.9 |
| Phonon description of temperature | 497.5 | 495.1 |

## 3. Results

### 3.1 Dislocation motion via a single kink pair in a small specimen

The CAC finite element model with dimensions of 201 nm x 123nm x 1008 nm is a relatively small one among Set 1 computer models listed in Table 1, comprising 2 billion atoms. Simulation results show that the dislocation in this small specimen moves through the process of nucleation and propagation of a single kink pair. In Fig. 11, we presents simulation result of this model for the dislocation motion visualized using the Dislocation Extraction Algorithm (DXA)[59]. As can be seen from Fig. 11(a), under an applied stress of 535 MPa, the dislocation remains in its initial Peierls valley. When the applied shear stress reaches 555 MPa, the dislocation starts to move through the formation of a kink pair, as shown in Fig. 11(b). This kink pair propagates along the dislocation line, as shown in Fig. 11(c), which causes the entire dislocation to translate to the next Peierls valley. Based on this observation, the critical stress required to initiate dislocation motion is expected to lie between 535 MPa and 555 MPa.

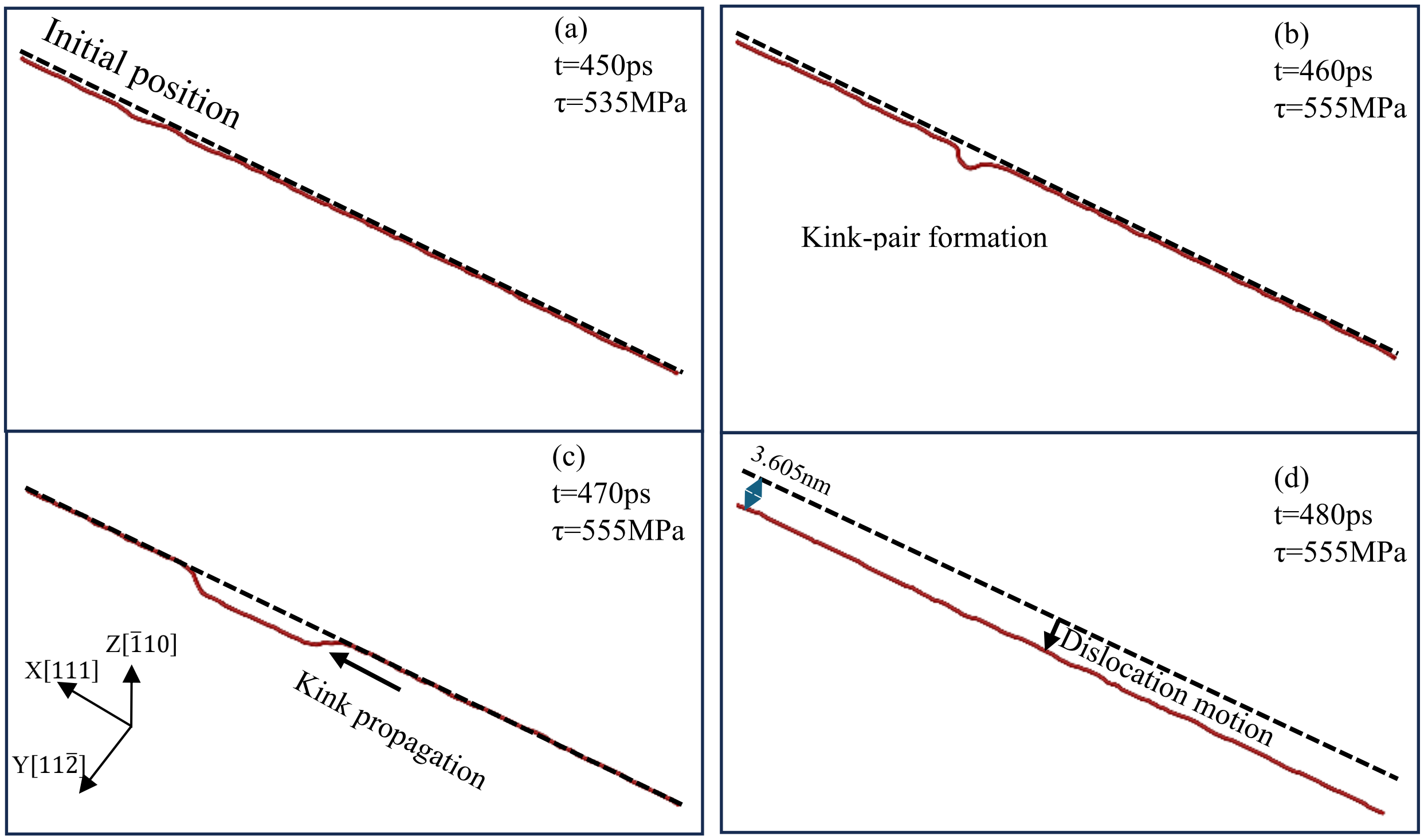

Fig. 11 Dislocation motion visualized by DXA from the specimen that has dimensions of 201 nm x 123 nm x 1008 nm and contains a dislocation with length of 201 nm, under incrementally increasing load from (a) 535MPa at t = 450 ps to (d) 555MPa at t = 485 ps. The red lines represent the positions of the ½<111> screw dislocation. The dashed lines represent the initial position of the dislocation before the applied stress.

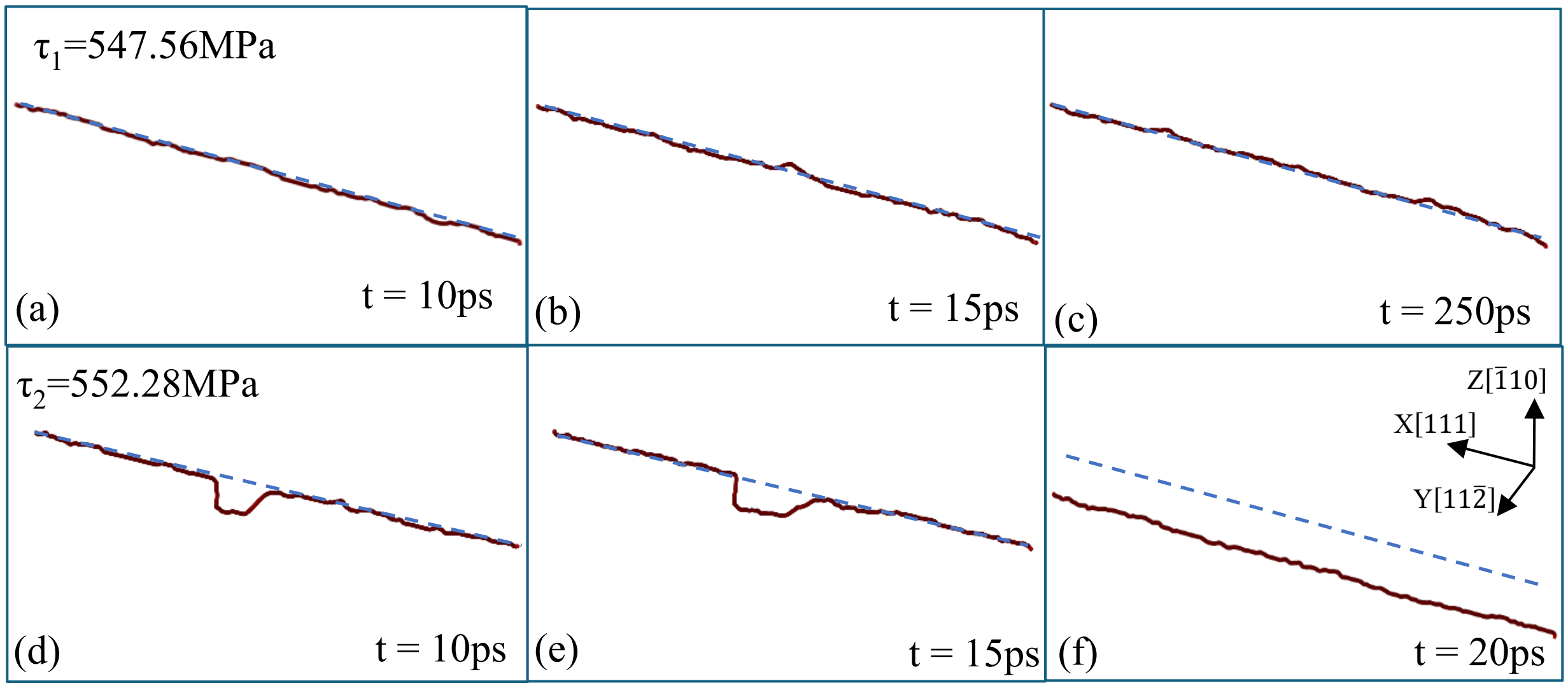

Fig. 12 Snapshots of dislocation motion visualized by DXA from the specimen that has dimensions of 201 nm x 123 nm x 1008 nm and contains a dislocation with length of 201 nm, under two different shear stress values, $\tau_1 = 547.56$ MPa and $\tau_2 = 552.28$ MPa. The red lines represent the ½<111> screw dislocation. The blue dashed lines represent the initial position of the dislocation before the application of the shear stress.

Based on this initial finding, five simulations were performed using respectively applied shear stress values of 538.12, 542.84, 547.56, 552.28, and 555.0 MPa. Fig. 12 presents snapshots that show the evolution of dislocation position in time. Under an applied shear stress of 547.56 MPa, the dislocation remained stationary throughout the 250 ps. In contrast, when the stress reached 552.28 MPa, the dislocation began to move continuously until it completely exited from its initial position. Therefore, the critical shear stress for this specimen was estimated as 552.28MPa, which is significantly lower than the reported value of 0.8-1.4 GPa from existing atomistic simulations in the literature.

### 3.2 Dislocation motion via multiple kink pairs in a large specimen

The CAC finite element model with dimensions of 2104 nm x 123 nm x 1008 nm is the largest computer model in Set 1, equivalently containing about 20.85 billion atoms. Figure 13 presents a zoomed-in view of the dislocation motion. As can be seen from Fig. 13 (a), under a constant shear stress of 408.40 MPa, at time t = 15 ps the dislocation has moved; significant dislocation migration with multiple kink pairs can be observed at t = 65 ps in zoomed-in Fig. 13(b). Formation of multiple kink pairs enables the dislocation to move across more than one Peierls valley before the entire line leaves its initial position, indicating that the applied stress has exceeded the critical stress for initiating dislocation motion.

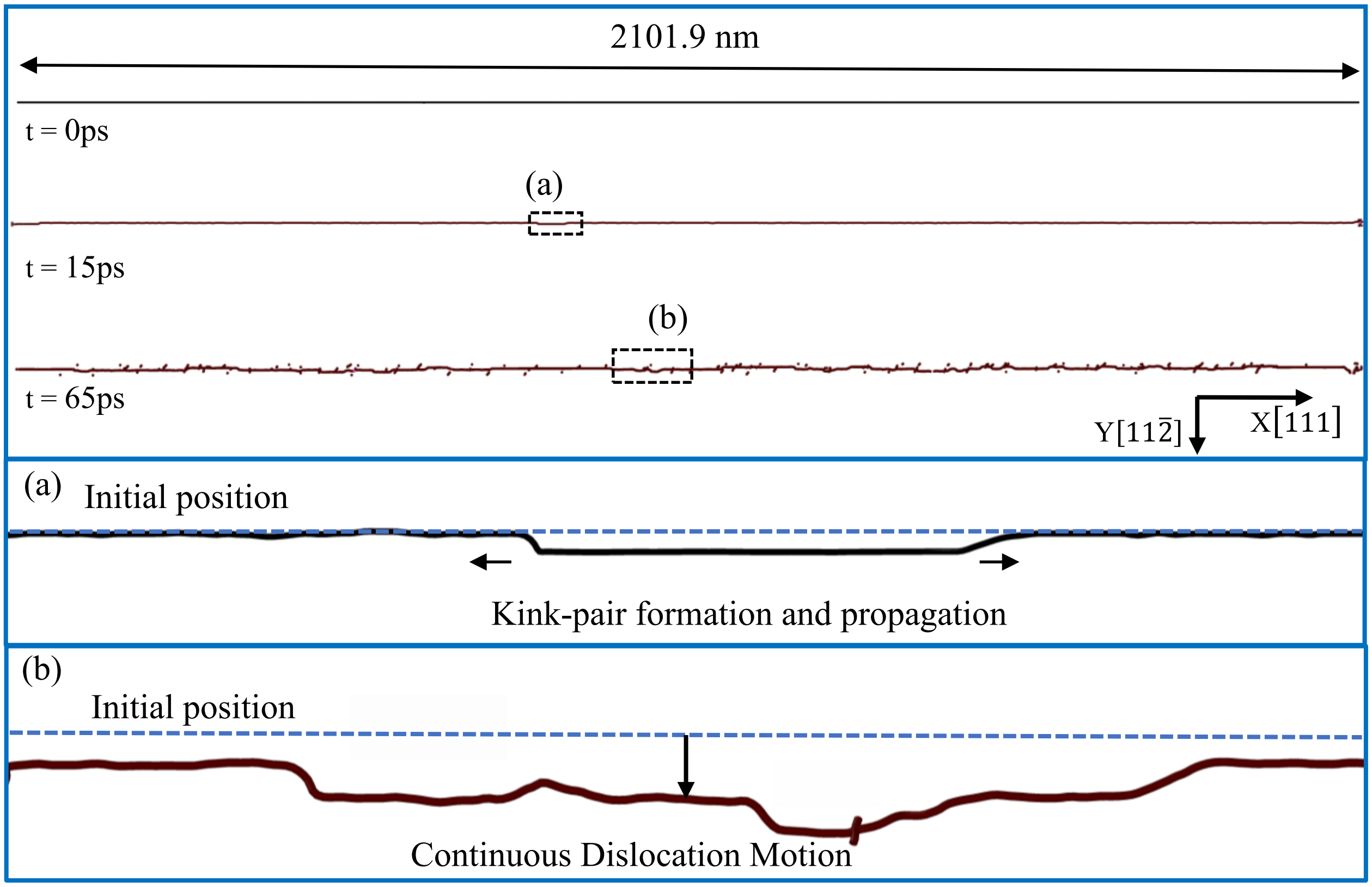

Fig. 13. Snapshots of ½<111> screw dislocation motion visualized by DXA from the specimen that has dimensions of 2104 nm x123 nm x1008 nm and contains a screw dislocation with length of 2.1 μm under a constant shear stress of 408.40 MPa; (a) and (b) present the zoomed-in view of the dislocation motion, in which the black lines represent dislocation line and the blue dashed lines represent the initial position of the dislocation.

### 3.3 Phonon-dislocation interaction

To visualize phonon-dislocation interaction, in Fig. 14 we present the time sequence of kinetic energy for the specimen with dimensions of 401 nm x123 nm x1008 nm under a constant shear stress of 497 MPa. It is seen from Fig.14 (a) that the interaction of phonons with the dislocation leads to an increase of kinetic energy, and subsequently the dislocation has moved to a new location shown in Fig. 14(c). Fig. 15 presents the time sequence of the displacement field in the (111) direction for the same specimen. As can be seen from Fig. 15, a strong interaction between the dislocation and the phonons leads to a large maximum displacement of 0.12 Å, whereas that for the perfect crystal presented in Fig. 6 is only 0.05 Å.

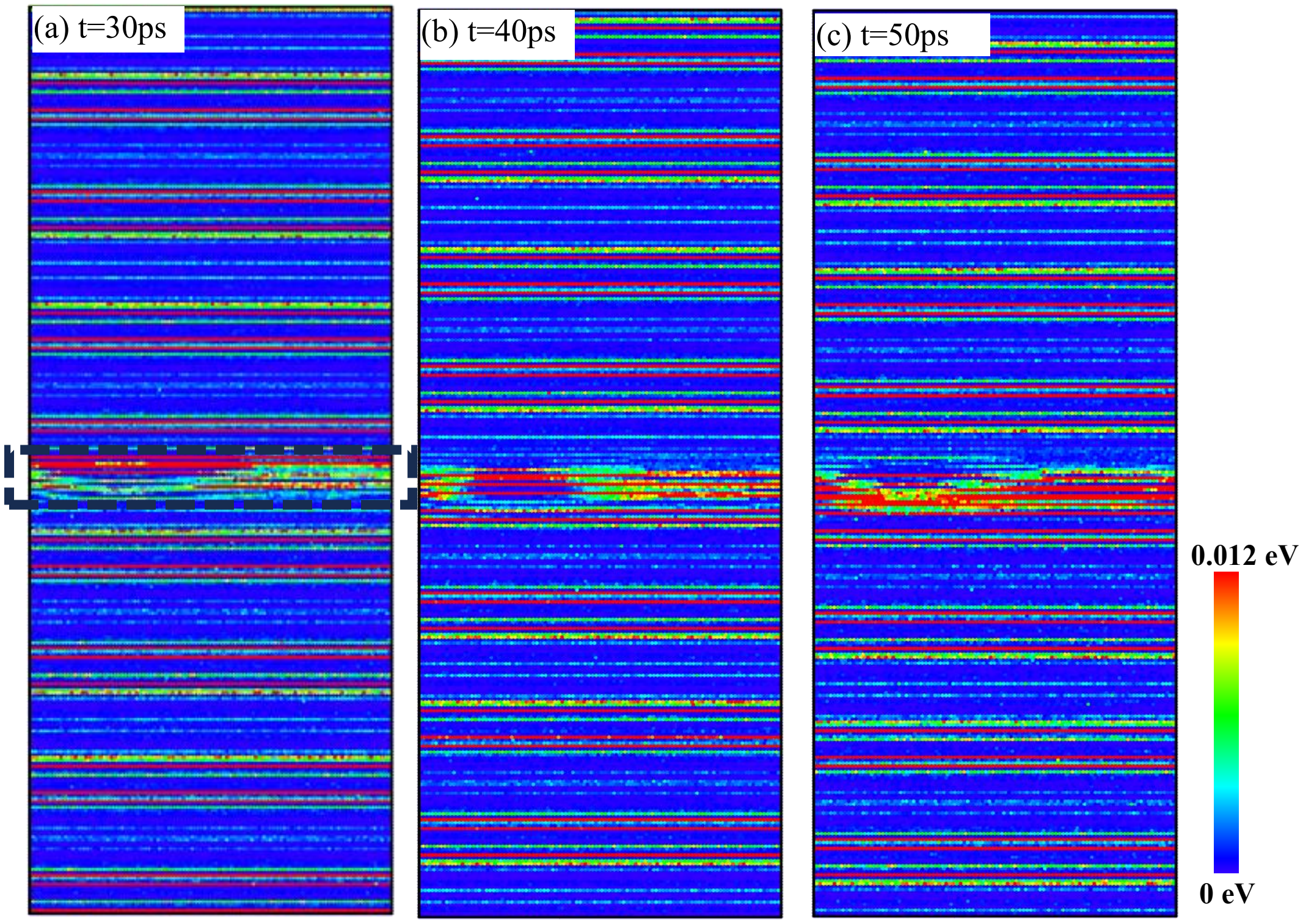

Fig. 14 Time sequence of the kinetic energy in the specimen that has dimensions of 401nm x123nm x1008nm and contains a screw dislocation. The dash-lined box in (a) marks the region that contains the dislocation and has the highest kinetic energy due to phonon-dislocation interaction.

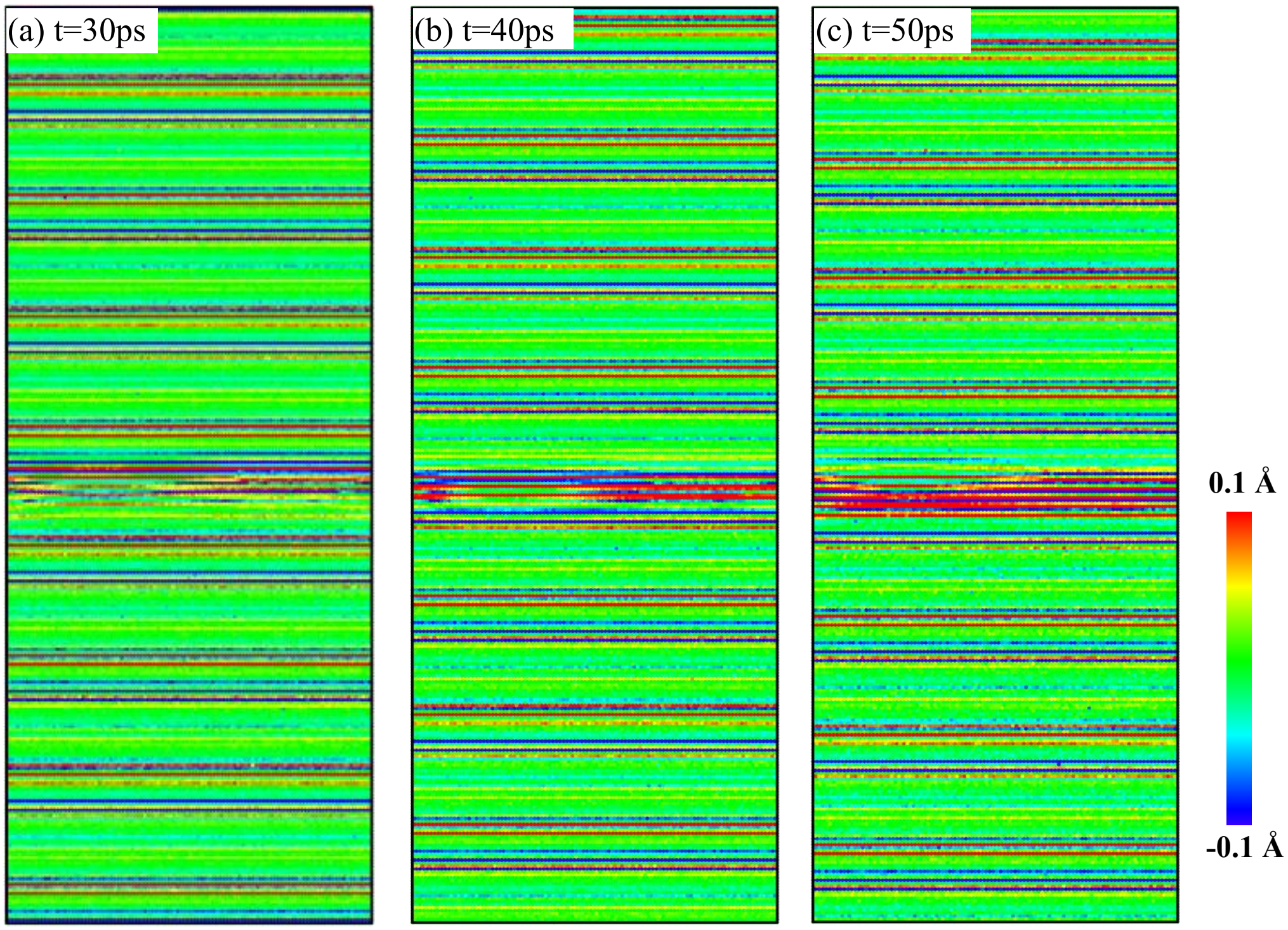

Fig. 15 Time sequence of the displacement field in the (111) direction within the specimen that has dimensions of 401 nm x123 nm x1008 nm.

Figure 16 presents the kinetic energy and displacement field in the 2101 nm × 123 nm × 1008 nm computer model. Larger kinetic energy can be observed to concentrate near the center of the model where the dislocation resides, indicating that phonon–dislocation interactions have occurred. To clarify these interactions, Fig. 16(b) provides a zoomed-in view of the kinetic energy and displacement field within rectangle A—the region enclosed by the dotted frame in the XZ plane of Fig. 16(a). This is accompanied by a zoomed-in DXA visualization of the dislocation segment, located in the XY plane of Fig. 16(a). A pronounced phonon–dislocation interaction is visible at the kink-pair locations, producing a dot-like pattern in the region of maximum displacement. This wave-interference pattern demonstrates that the incident phonon waves are reflected by the dislocation.

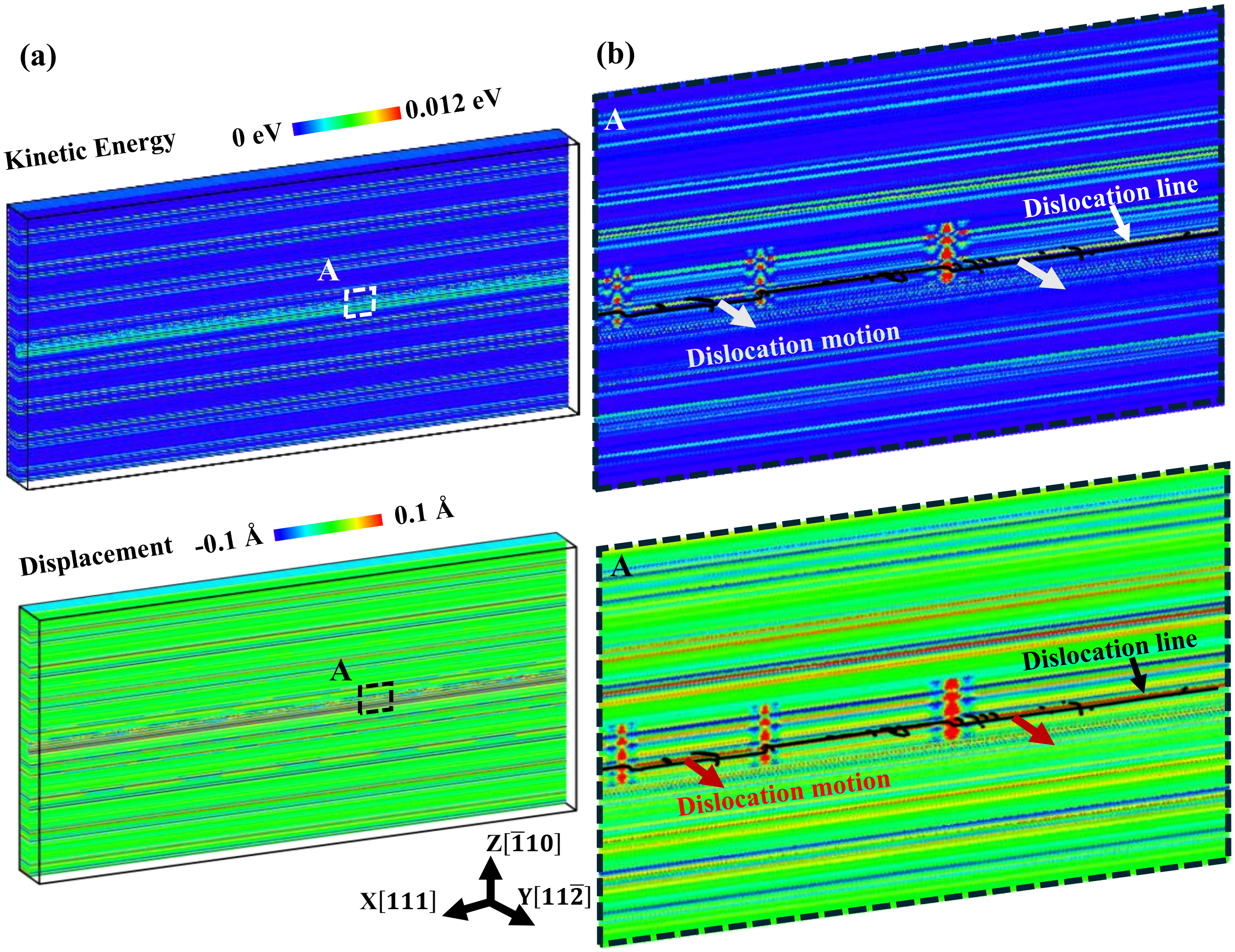

Fig. 16. A 3D view of the kinetic energy and displacement field in the 2101 nm × 123 nm × 1008 nm computer model; (b) Zoomed-in view of rectangle A in (a), showing strong wave reflections by the dislocation. Panel (b) also includes the DXA visualization of the dislocation segment (black); here, the solid black line represents the dislocation line, and the arrow indicates the direction of dislocation motion.

To quantify the effect of the transverse acoustic (TA) in phonon-dislocation interaction, we performed CAC simulations with the heat source that consists of TA phonons only and that has combined TA + LA phonons, respectively. The corresponding Peierls stresses were both measured to be 497.51 MPa, indicating that the TA phonons predominantly determine the Peierls stress. This behavior is consistent with the phonon polarizations in the present model, where TA phonons vibrate along the dislocation slip plane, while LA phonons vibrate perpendicular to the slip plane.

### 3.4 Effects of dislocation line length and definitions of temperature on Peierls stress

CAC simulation results for the five computer models described in Set 1 of Table 1 are summarized in Table 4 and plotted in Fig. 17. It is seen from Table 3 and Fig. 17 that:

(1) The computed Peierls stress is significantly smaller in simulations using the phonon definition of temperature than those using the classical kinetic temperature. Compared with the experimental value of ~ 375 MPa at 0.7K, the discrepancies (values in parentheses) for simulations using the classical temperature range from 223-251%, whereas those for the phonon definition of temperature range from 8.9-56.8%, demonstrating a significant difference in the computed Peierls stress caused by different temperature definitions.

(2) The computed Peierls stress decreases as the dislocation line length increases. With the kinetic temperature definition, the Peierls stress varies slightly from 1318 MPa for dislocation with length of 100 nm to 1210 MPa for dislocation length of 2101 nm. The overall reduction in Peierls stress is approximately 8%, showing a limited size effect. In contrast, using the phonon description of temperature, the Peierls stress decreases from 588 MPa for dislocation with line length of 100 nm to 408 MPa for dislocation with line length of 2102 nm, indicating a stronger size effect caused by phonon-dislocation interactions.

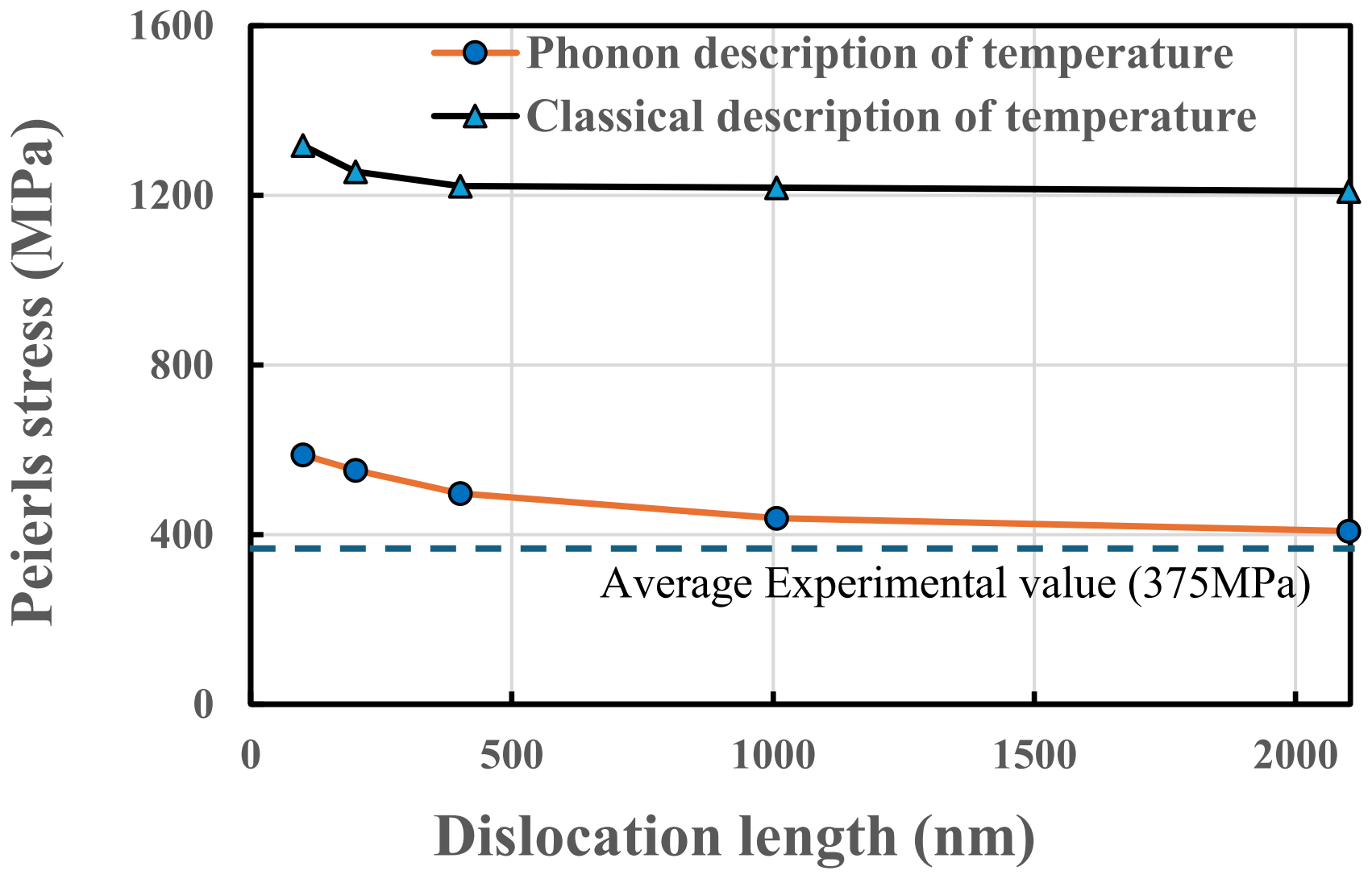

Fig. 17 CAC simulation results of Peierls stress as a function of dislocation length, showing the significant effect of temperature definition. All specimens have fixed dimensions of 123 nm in the Y direction and 1008 nm in the Z direction, with the length ranging from 100 nm to 2102 nm. The experimental value 375 MPa is based on averaging results in the literature[1-3,60] as indicated by the blue dashed line.

Table 3. Computed Peierls stress (MPa) as a function of dislocation length at a temperature of 0.7K, using classical and phonon description of temperature, respectively. All specimen has thickness of 123nm. The values in parentheses indicate the discrepancies with the experimental value of 375MPa.

| Dislocation line length (nm): | 99.7 | 200.6 | 400.6 | 1005.9 | 2101.9 |
|---|---|---|---|---|---|
| Peierls stress (MPa) obtained using Classical temperature | 1317.766 (251.40%) | 1256.348 (235.03%) | 1222.329 (225.95%) | 1217.647 (224.71%) | 1210.232 (222.73%) |
| Peierls stress (MPa) obtained using Phonon temperature | 588.23 (56.86%) | 552.28 (47.27%) | 497.51 (32.67%) | 438.82 (17.02%) | 408.40 (8.915%) |

### 3.5 Effects of the specimen thickness

The goal of the simulations of the models for Set 2 is to quantify the effect of specimen thickness. Fig.18 presents simulation results for the Peierls stress as a function of specimen thickness, i.e., the dimension in the $[11\bar{2}]$ direction. All the specimens contain a 400 nm-long dislocation, with thickness ranging from 41 nm to 491 nm. It is seen from Fig.18 that the Peierls stress decreases from 513.2 MPa for specimen with thickness of 41 nm to 497.5 MPa for a thickness of 123 nm, a reduction of 3.2%. However, the Peierls stress appears to converge as the thickness increases, since it decreases only slightly from 496.13 MPa for specimen thickness of 245 nm to 495.40 MPa for specimen thickness of 491 nm.

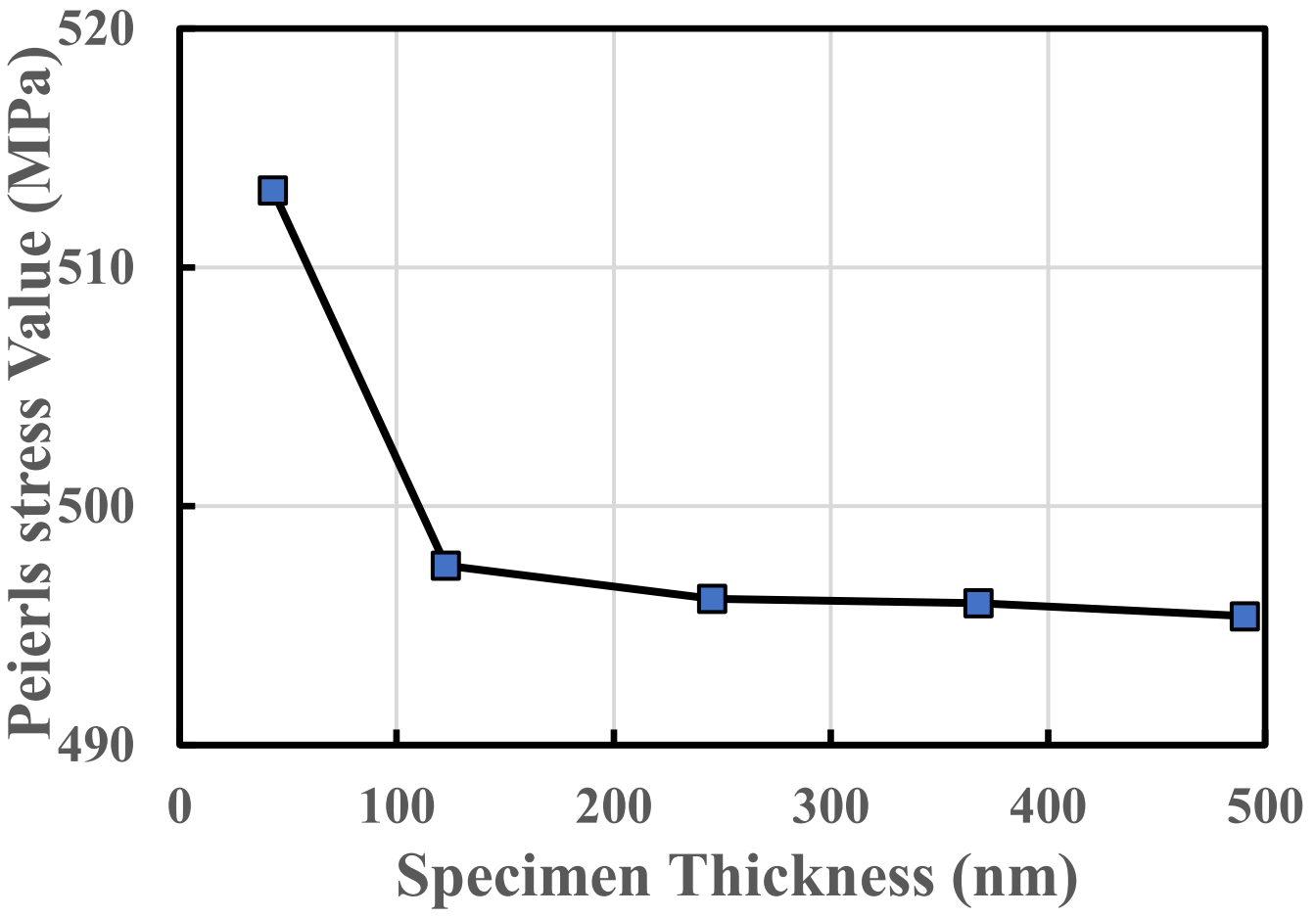

Fig. 18 Peierls stress as a function of specimen thickness. All specimens contain a 400 nm-long dislocation, with thickness ranging from 41 nm to 491 nm. All simulations use the phonon description of temperature.

To further confirm the effect of specimen thickness, Fig. 19 presents simulation results of three large specimens, each containing a 2101.9 nm-long dislocation, with thicknesses of 41 nm, 123 nm, and 245 nm, respectively. As can be seen from Fig.19, as the specimen thickness increases from 123 nm to 245 nm, the Peierls stress decreases from 408.40 MPa to 408.11 MPa, a 0.07% decrease.

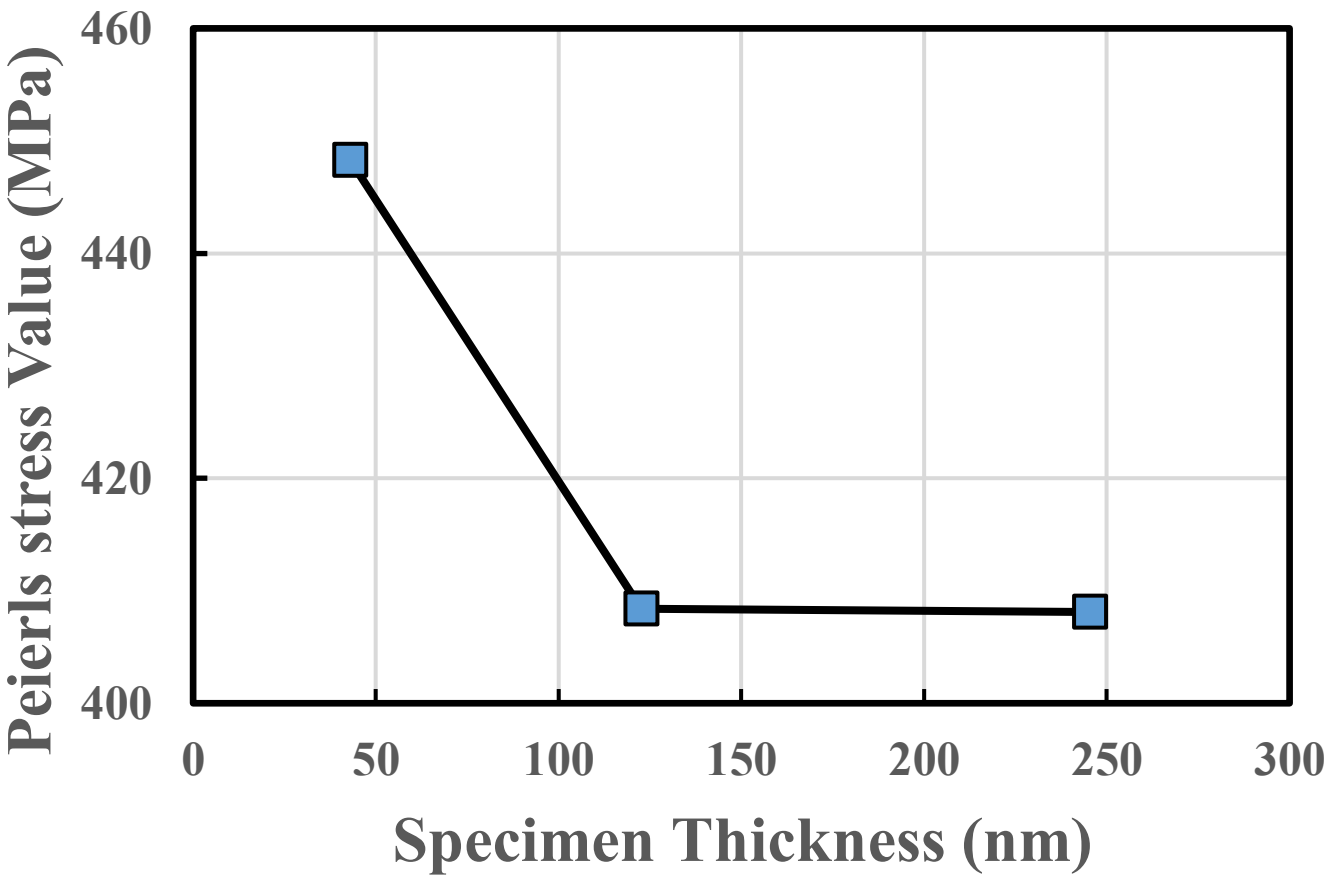

Fig. 19 Peierls stress versus specimen thickness ($[11\bar{2}]$ direction). All specimens contain a 2101.9 nm-long dislocation, with thickness ranging from 41 nm to 245 nm. All simulations use the phonon description of temperature.

## 4 Summary and Conclusions

In summary, we have presented a concurrent atomistic-conitnuum (CAC) method with phonon description of temperature to preserve the Bose-Einstein distribution of phonon modes at low temperature and to reproduce the nucleation and propagation of kink-pairs with dislocation line length ranging from 0.1 to 2.1μm. The methodology has been shown to be able to reproduce a constant average kinetic energy and a constant average displacement field at a low temperature of 0.7K. Its prediction of the Peierls stress is comparable to that of MD simulations for small-scale specimens using the kinetic temperature definition with significantly fewer DOFs. We have presented two sets of CAC simulation results of Peierls stress for 1/2 < 111 > screw dislocation lines in BCC iron, with different specimen sizes, dislocation line lengths, and temperature definitions. Major findings of this work are summarized as follows.

(1) The quantum effect on Peierls stress is striking, manifested as an approximately three times difference in measured Peierls stress between simulations using classical temperature and the phonon description of temperature. This essentially explains much lower experimentally measured values of the Peierls stress than those resulting from DFT calculations and MD simulations.
(2) The dependence of the Peierls on dislocation line length has been quantified. Using the phonon definition of temperature, the value of Peierls stress has been shown to decrease as the dislocation line length increases from 99.7 nm to 2101.9 nm. The smallest Peierls stress, 408.11 MPa, was obtained for the longest dislocation in the largest specimen with dimensions 2104 nm x 245nm x 1008 nm, which would contain 41.7 billion atoms if simulated by MD. This value, if extrapolated to zero temperature, should be very close to the experimentally estimated Peierls stress of 375 MPa.
(3) The effect of specimen size on Peierls stress has also been quantified. CAC simulation results based on the phonon description of temperature show that the value of the Peierls stress decreases as the specimen length or thickness increases. It is noted that the Peierls stress computed in this work has nearly converged to that of experimentally measured values as the dimensions of the specimen increase to 2104 nm x 245 nm x 1008 nm.
(4) Short wavelength phonons have been shown to have negligible effect on the motion of a dislocation with long line length at a low temperature. This indicates that the strength of interaction between phonons and dislocations therefore depend on dislocation length and phonon wavelength.

Finally, it may be worth noting that this work used only a single 1/2 < 111> pure screw dislocation. Future work may quantify the effect of collective motion of many dislocations on the Peierls stress, including different mixed-type dislocations.

### Acknowledgement

This work was supported by the U.S. National Science Foundation (NSF) under Award Number CMMI-2054607. Chen and Li were also partially supported by the NSF under Award Numbers 2349160 and 2121895, respectively. Computational resources were provided by the Advanced Cyberinfrastructure Coordination Ecosystem: Services & Support (ACCESS) under allocation TG-DMR190008.